\documentclass[twocolumn,prd,nofootinbib,aps,floats,floatfix,amsmath,amssymb,secnumarabic]{revtex4}

\usepackage{amsmath}
\usepackage{verbatim}
\usepackage{graphicx}
\usepackage[usenames]{color}
\usepackage{psfrag}
\usepackage{epstopdf}

\newbox\tablebox    \newdimen\tablewidth
\def\leaderfil{\leaders\hbox to 5pt{\hss.\hss}\hfil}
\def\endtable{\tablewidth=\columnwidth 
    $$\hss\copy\tablebox\hss$$
    \vskip-\lastskip\vskip -2pt}

\def\tablenote#1 #2\par{\begingroup \parindent=0.8em
    \abovedisplayshortskip=0pt\belowdisplayshortskip=0pt
    \noindent
    $$\hss\vbox{\hsize\tablewidth \hangindent=\parindent \hangafter=1 \noindent
    \hbox to \parindent{\sup{\rm #1}\hss}\strut#2\strut\par}\hss$$
    \endgroup}
\def\doubleline{\vskip 3pt\hrule \vskip 1.5pt \hrule \vskip 5pt}

\begin{document}

\title{Minimal parameterizations for modified gravity}

\author{Ali Narimani} 
\email{anariman@phas.ubc.ca}
\affiliation{Department of Physics \& Astronomy,\\
University of British Columbia,\\
6224 Agricultural Road,\\
Vancouver, BC, V6T 1Z1  Canada}
\author{Douglas Scott} 
\email{dscott@phas.ubc.ca}
\affiliation{Department of Physics \& Astronomy,\\
University of British Columbia,\\
6224 Agricultural Road,\\
Vancouver, BC, V6T 1Z1  Canada}

\begin{abstract}

The increasing precision of cosmological data provides us with an opportunity to test general relativity (GR) on the largest accessible scales.
Parameterizing modified gravity models facilitates the systematic testing of the predictions of GR, and gives a framework for detecting possible deviations from it.
Several different parameterizations have already been suggested, some
linked to classifications of theories, and others more empirically motivated.
Here we describe a particular new approach which casts
modifications to gravity through two free functions of time and scale,
which are directly linked to the field equations, but also easy to confront with observational data.
We compare our approach with other existing methods of parameterizing modied gravity, specifically the parameterized post-Friedmann approach and the older method using the parameter set $\{\mu,\gamma\}$.
We explain the connection between our parameters and the physics
that is most important for generating cosmic microwave background anisotropies.
Some qualitative features of this new parameterization, and therefore modifications to the gravitational equations of motion, are illustrated in a toy model, where the two functions are simply assumed to be constant parameters.

\end{abstract}

\maketitle
\newpage
\section{Introduction}
General Relativity (GR) has been confronted with many theoretical and experimental tests since 
its birth in $1915$. From the gravitational lensing experiments in $1919$ \cite{eddin} up to the 
extensive studies and tests in the $1960$s and $1970$s 
\cite{shapiro,kerr,penrose,hawking,taylor}, the theory has been confirmed observationally and
theoretically bolstered in many different respects. Weak field gravity on experimentally 
accessible scales has been so well tested that there are only two remaining directions in 
which we might find modifications to GR: strong field gravity, which may be probed by studying 
black holes; and gravity at large scales and early times, which is the cosmological arena.

Cosmology has challenged GR with two, yet to be fully understood discoveries:
dark matter and dark energy \cite{DM,DE}. 
Along with these two phenomena, the lack of renormalizability in GR \cite{clifton} and the 
apparently exponential expansion in the very early Universe \cite{guth} 
are usually taken as signs for the incompleteness of the theory at high energies.
Due to these shortcomings in GR the study of modified gravity has become a broad field.
Scalar-tensor theories \cite{brans,nord}, $f(R)$ modifications \cite{nojiri,faraoni}, 
Horava-Lifshitz theory \cite{horava}, multidimensional theories of gravity 
\cite{kolb,seto}, and many other suggestions have been made in the hope of finding, or at 
least deriving, some hints for, a fully consistent theory that can successfully explain 
the observations and satisfy the theoretical expectations. (Ref.~\cite{clifton} has an 
extensive review).

The new data coming from various experiments such as the \textit{WMAP} and \textit{Planck} satellite
measurements of the cosmic microwave background (CMB) anisotropies \cite{tauber}, and the WiggleZ 
\cite{wigglez} or Baryon Oscillation Spectroscopic Survey \cite{boss} measurements 
of the matter power spectrum, provide us with an opportunity to test specific modified theories
of gravity. However, since there are many different modified theories, all with their own sets of
parameters, there has recently been some effort to come up with a way to describe
generic modified theories using only a few parameters, and to try to constrain those parameters
with general theoretical arguments and by direct comparison with cosmological data.

Parametrizing modified theories of gravity with a small number of parameters has the benefit of
tracking the effects of modified gravity on a number of different observables consistently and
systematically, rather than considering the consistency of each single observable with GR 
predictions. This can,
at least in principle, lead to constraints on the theory space of 
modified gravity models.

The parametrized post-Friedmann (PPF) approach, as described in Ref.~\cite{baker}, is an effective
way to parameterize many of the modified theories of gravity. However, it is not
really feasible to constrain its more or less dozen additional free functions, even with the power of
Markov Chain codes such as 
CosmoMC \cite{cosmomc}; there are just too many degrees of freedom to provide useful constraints
in the general case. 
In this paper we will describe a somewhat different way to parametrize 
modified theories of gravity in which we try to retain only a small number of 
parameters, which we then constrain using \textit{WMAP} $9$-year \cite{WMAP9} and
SPT data \cite{SPT12}.

In the next section we will describe the formulation of this new parametrization, and
will show its 
connection with PPF and other approaches in Sec.~\ref{relation}. In Sec.~\ref{numeric} we will 
discuss the results of a numerical analysis using CAMB \cite{CAMB} and CosmoMC, and we 
will conclude the paper in Sec.~\ref{discuss} with a brief discussion.

\section{Modified gravity formulation}
\label{ABparam}
There are two common strategies for modifying gravity. One can start 
from the point of view of the Lagrangian or from the equations of motion. The 
Lagrangian seems like the more obvious path for writing down specific new theories, 
where one imagines retaining some desired symmetries while breaking 
some others. However, the equations of motion provide an easier way in practice to parametrize 
a general theory of modified gravity, especially in the case of first order perturbations in 
a cosmological context. 

The evolution of the cosmological background has been well tested at  
different redshift slices, specifically at Big Bang nucleosynthesis and at
recombination through the CMB anisotropies. It therefore seems reasonable to assume that 
the background evolution is not affected by the gravity modification, with the only 
background level effect being a possible explanation for a fluid behaving like dark energy. 

The linearized and modified equations of motion for gravity can 
be written in the following form in a covariant theory:
\begin{equation} \label{eins}
\delta G_{\mu \nu} = 8 \pi G \, \delta T_{\mu \nu} + \delta U_{\mu \nu}.
\end{equation}
Here, $\delta G_{\mu \nu}$ is the perturbed Einstein tensor around a background metric,
$\delta T_{\mu \nu}$ is the first order perturbation in the energy-momentum tensor and 
$\delta U_{\mu \nu}$ is the modification tensor source from any term that is not 
already embedded in GR. 


Since we will be using CAMB for numerical calculations, 
we will choose the synchronous gauge from now on, and focus only on the spin-$0$ (scalar) 
perturbations. 
This will make it much more straightforward to adapt the relevant perturbed Boltzmann 
equations. The metric in the synchronous gauge is written as
\begin{eqnarray}
&&ds^2 =  a^2(\tau) \left[ -d\tau^2 + ( \delta_{ij} + h_{ij} ) dx^idx^j \right] , \nonumber   \\
&&h_{ij} =  \int d^3k e^{i\vec{k}.\vec{x}} \lbrace \hat{k}_i \hat{k}_j h(\vec{k},\tau)
+(\hat{k}_i \hat{k}_j-\dfrac{1}{3} \delta_{ij})6 \eta(\vec{k},\tau) \rbrace , \nonumber \\
&& \vec{k}  = k\hat{k} ,
\end{eqnarray}
where $\vec{k}$ is the wave vector. Putting this metric into Eq. \ref{eins} results 
in the following four equations
\cite{ma}:
\begin{eqnarray}
&&k^2  \eta - \frac{1}{2} \, \frac{\dot{a}}{a} \, \dot{h} =
- 4 \, \pi \, G \, a^2 \delta \rho + k^2 \, A(k,\tau) ; \label{first}  \\
&&k  \dot{\eta} = 4 \, \pi \, G \, a^2 (\bar{\rho}+\bar{p}) \,  V 
+ k^2  B(k,\tau) ; \label{second} \\
&&\ddot{h} + 2\,\frac{\dot{a}}{a} \, \dot{h} - 
2\, k^2 \, \eta = - 24\, \pi \, G \, a^2 (\delta P) 
+ k^2\, C(k,\tau) ; \label{third} \\ 
&&\ddot{h} + 6 \ddot{\eta}  + 2 \frac{\dot{a}}{a} 
(\dot{h} + 6 \dot{\eta}) - 2\, k^2 \eta =
-24 \pi  G \, a^2 (\bar{\rho}+\bar{p})\Sigma \label{last}
 \nonumber \\
 && + k^2 D(k,\tau)  .
\end{eqnarray}
Here we have used the following definitions:
\begin{equation}
  \begin{split}
    \delta T^0_0 &= -\delta \rho \\
    \delta T^0_i &= (\bar{\rho}+\bar{p}) \,V_i \\
    \delta T^i_i &= 3\,\delta P  \\
    \mathcal{D}_{ij} \delta T^{ij} & =  (\bar{\rho}+\bar{p})\Sigma
  \end{split}
\quad \begin{split}
 &, \\
 &, \\
 &, \\
 &,
 \end{split}\quad
  \begin{split}
    a^2\delta U^0_0 &= k^2\,A(k,\tau),\\
    a^2\delta U^0_i &= k^2\,B(k,\tau),\\
    a^2\delta U^i_i &= k^2\,C(k,\tau),\\
    a^2 \mathcal{D}_{ij} \delta U^{ij}  &=  k^2\,D(k,\tau) , \label{param}
  \end{split}
\end{equation}
with $\bar{\rho}$ and $\bar{p}$ being the background energy density and pressure, 
respectively, and a dot representing a derivative with respect to $\tau$. The factors 
of $k$ are chosen to make the modifying functions, $\{A,B,C,D\}$, dimensionless. The quantity
$\mathcal{D}_{ij}$ is defined as $\hat{k}_i \hat{k}_j-\dfrac{1}{3} \delta_{ij}$.
The parameterization described here has a very close connection in practice with
the PPF method explained in Ref.~\cite{baker}. The most important differences are that we have grouped a
number of separate parameters into a single parameter, and have used the synchronous gauge in 
Eqs.~\ref{first} to \ref{last}.

In general, Einstein's equations provide six independent equations. For the case of
first order perturbations in cosmology, two of these six equations are used for the 
two spin-2 (tensor) degrees of freedom, two of the equations are used for the spin-1 (vector)
variables and only two independent equations are left for the spin-0 (scalar) degrees of freedom.
This means that Eqs.~\ref{first} to \ref{last} are not independent and one has to 
impose two consistency relations on this set of 
four equations. These consistency relations of course come from the energy-momentum 
conservation equation, $\nabla_\mu (T^\mu_{\,\,\, \nu} + U^\mu_{\, \,\, \nu}) =0 $.
Assuming that energy conservation holds independently for the conventional fluids,
$\nabla_\mu T^\mu_{\,\,\, \nu} =0 $, (see Ref.~\cite{baker} for the strengths and weaknesses 
of such an assumption) one then obtains the following two consistency equations:
\begin{eqnarray}
\dfrac{2\,\dot{A}}{\mathcal{H}}
+2\,A-\dfrac{2\,k\,B}{\mathcal{H}}+C&=&0 \,  ; \label{yek} \\
6\,\dot{B}+k\,C+12\,\mathcal{H}\,B-k\,D&=&0 \, \label{do}.
\end{eqnarray}
Here we have defined $\mathcal{H} \equiv \dot{a}/a$ and dropped the arguments of the functions
$A$ to $D$. 

Eqs.~\ref{first} to \ref{last}, together with Eqs.~\ref{yek} and \ref{do}, show
that two general functions of space and time would be enough to parametrize a wide range 
of modified theories of gravity. This approach of course does not provide a test for any
specific modified theory. However, given the current prejudice that GR \textit{is} the 
true theory of gravity at low energies (e.g. see Ref.~\cite{burgess} for a discussion), the main 
question is whether or not cosmological data 
can distinguish between GR and any other generic theory of modified gravity.
Clearly, if we found evidence for deviations from GR, then we would have a parametric way of 
constraining the space of allowed models, and hence hone in on the correct theory.

\section{Connection with other methods of parametrization}
\label{relation}
In this section we show the connection between
the conventional $\{\mu,\gamma\}$ parameterization of modified gravity,
the PPF parameters and the parameterization introduced in 
Sec.~\ref{ABparam}.

The parameters defined in Sec.~\ref{ABparam}
are related to the PPF parameters according to
\begin{eqnarray}
A(k,\tau) &=& A_0 \hat\Phi+F_0\hat\Gamma+\alpha_0\hat\chi+\frac{\alpha_1}{k}\dot{\hat\chi}, \label{bir} \\
B(k,\tau) &=& B_0 \hat\Phi+I_0\hat\Gamma+\beta_0 \hat\chi+\frac{\beta_1}{k}\dot{\hat\chi} , \\
C(k,\tau) &=& C_0 \hat\Phi+ \frac{C_1}{k} \dot{\hat\Phi}+J_0\hat\Gamma+\frac{J_1}{k} \dot{\hat\Gamma}
+\gamma_0 \hat\chi+\frac{\gamma_1}{k}\dot{\hat\chi} \nonumber \\
&&+\frac{\gamma_2}{k^2} \ddot{\hat\chi}, \\
D(k,\tau) &=& D_0\hat\Phi+\frac{D_1}{k} \dot{\hat\Phi}+K_0\hat\Gamma+\frac{K_1}{k}\dot{\hat\Gamma}+
\epsilon_0\hat\chi+\frac{\epsilon_1}{k}\dot{\hat\chi}\nonumber \\
&&+\frac{\epsilon_2}{k^2} \ddot{\hat\chi} \label{dor} .
\end{eqnarray}
While the authors of Ref.~\cite{baker} have insisted on the modifications being gauge invariant, it is 
good to keep in mind that there is nothing special about the use of gauge invariant parameters, as is shown
in Ref.~\cite{malik}. The important issue is to track the degrees of freedom in the equations. There are 
originally four free functions for the spin-$0$ degrees of freedom in the metric, but the gauge freedom 
can be used to set two of them to zero. Using only 
two gauge invariant functions instead of four, means that the gauge freedom has been implicitly used 
somewhere to omit the redundant variables. 

All $22$ of the parameters on the right hand side of Eqs.~\ref{bir} to \ref{dor}
are in fact two-dimensional functions of 
the wave number, $k$, and time. A hat on a function means that it is a gauge invariant quantity. 
The symbol $\hat\chi$ is the gauge invariant form of any extra degree of freedom 
that can appear, for example, in a scalar-tensor theory, or in an $f(R)$ theory as a result of a 
number of conformal transformations (see section D.2 of Ref.~\cite{baker} for further explanation). 
$\hat\Phi$ and $\hat\Gamma$ are related to the synchronous gauge metric perturbations through:
\begin{eqnarray}
\hat{\Phi} &=& \eta - \frac{\mathcal{H}}{2\, k^2}(\dot{h}+6\, \dot{\eta});  \\
\hat{\Psi} &=& \frac{1}{2\,k^2} ( \ddot{h} + 6\, \ddot{\eta} + \mathcal{H} \, 
(\dot{h} + 6\, \dot{\eta} ) );  \\
\hat\Gamma &=& \frac{1}{k}(\dot{\hat{\Phi}}+\mathcal{H}\,\hat{\Psi} ) .
\end{eqnarray}
One needs to add more parameters to the right hand side of Eqs.~\ref{bir} to \ref{dor} 
if there is more than one extra degree 
of freedom, or if the equations of motion of the theory are higher than second order and the theory
cannot be conformally transformed into a second order theory. 
The reason this many parameters were introduced in Ref.~\cite{baker} is that
there is a direct connection between these parameters and the Lagrangians of a number of specific
theories, like the Horava-Lifshitz, scalar-tensor or Einstein Aether theories.
Therefore, in principle, constraining these parameters is equivalent to constraining 
the theory space of those Lagrangians. 

However, there are a number of issues that may encourage one to consider alternatives to the PPF  
approach for parameterizing modifications to gravity. First of all, it is practically impossible to 
run a Markov chain code for $22$ two-dimensional functions. One can reduce the number of free functions to
perhaps $15$ using Eqs.~\ref{yek} and \ref{do}, but there is still a huge amount of freedom in the problem.
The second reason is that the whole power of the PPF method lies in distinguishing among a number of
classically modified theories of gravity that are mostly proven to be either theoretically inconsistent,
like the Horava-Lifshitz theory \cite{clifton}, or already ruled out observationally, like TeVeS
(at least for explaining away dark matter) \cite{TeVeS}. 
While it is certainly important and useful to check the GR predictions with the new coming 
data sets, it does not appear reasonable at this stage to stick with the motivation of any specific theory.
For the moment it therefore seems prudent to consider an even simpler approach, as we describe here.

There is another popular parametrization in the literature, described fully in 
Refs.~\cite{alireza,levon1,levon2}. This second parametrization is best 
described in the conformal Newtonian gauge, via the following metric:
\begin{equation}
ds^2 = a^2(\tau)[-(1+2 \, \psi) d\tau^2 + (1 - 2 \, \phi)\delta_{ij}dx^idx^j] .
\end{equation}
The modifying parameters, $\{\mu,\gamma\}$, are defined through the following:
\begin{eqnarray}
&&k^2\psi = - \mu(k,a) 4 \pi G a^2 \lbrace \bar{\rho} \Delta + 3(\bar{\rho} + \bar{p}) 
\Sigma \rbrace  \label{mg-poisson} \, ; \\
&&k^2[\phi - \gamma(k,a) \psi] = \mu(k,a)  12 \pi G a^2   (\bar{\rho} + \bar{p})
 \Sigma \label{mg-anisotropy} \, .
\end{eqnarray}
Here $\Delta = \delta \rho + 3 \frac{\mathcal{H}}{k}(1 + \bar{p}/\bar{\rho}) V $, and all of the 
matter perturbation quantities are in the Newtonian gauge. 

In order to see the connection between this method of parametrization and the one described in 
the previous section through Eqs.~\ref{first} to \ref{last}, one needs to use the modified
equations of motion in the Newtonian gauge:
\begin{eqnarray}
&&k^2\phi + 3\mathcal{H} \left( \dot{\phi} + \mathcal{H}\psi
	\right) = - 4\pi G a^2 \delta \rho + k^2 A_{\rm N} \,;
	\label{ein-cona}\\
&&k^2 \left( \dot{\phi} + \mathcal{H}\psi \right)
	 = 4\pi G a^2 (\bar{\rho}+\bar{p}) k\,V  + k^3 B_{\rm N}
	 \,;\label{ein-conb}\\
&&k^2(\phi-\psi) = 12\pi G a^2 (\bar{\rho}+\bar{P})\Sigma  +k^2 D_{\rm N}
	\,.\label{ein-cond}
\end{eqnarray}
The parameters $\{A_{\rm N},B_{\rm N},D_{\rm N}\}$ are the modifying functions in the Newtonian gauge. 
These parameters are related to $\gamma$ and $\mu$ via
\begin{eqnarray}
&&\alpha \equiv  1-\mu \, ,    \\
&&\beta  \equiv  \gamma -1 ,  \\
&&4 \pi G a^2 \alpha \lbrace \bar{\rho} \Delta + 3(\bar{\rho} + \bar{p}) \Sigma \rbrace =
k^2(A_{\rm N}-3\frac{\mathcal{H}}{k}B_{\rm N}  \nonumber  \\
&& \qquad + D_{\rm N} ) ,  \\
&&\beta \, \psi = 12\, \pi G a^2 \, \alpha \Sigma + k^2 \, D_{\rm N} ,
\label{Agamma}
\end{eqnarray}
where one can choose between using the functions $\{A_{\rm N},B_{\rm N},C_{\rm N},D_{\rm N}\}$, along 
with two constraint equations 
similar to the Eqs.~\ref{yek} and \ref{do}, or using the two parameters $\gamma$ and $\mu$
and trying to remain consistent in the equations of motion. 

It is argued in Ref.~\cite{tessa} that the $\{\gamma,\mu\}$ choice is not capable 
of parameterizing second order theories in the case of an unmodified background and no 
extra fields. To show this the authors use the fact that, in the absence of extra fields, 
all of the Greek coefficients in Eqs.~\ref{bir} to \ref{dor}, 
i.e. $\{\alpha_0,..., \epsilon_2\}$, have to be zero. Furthermore, they argue that in the 
case of second order theories, $F_0$ and $I_0$ have to be zero, 
and therefore the constraints of Eqs.~\ref{yek}
and \ref{do} show that $J_0$ and $K_0$ are zero as well. After all of this, one can see that 
Eq.~\ref{ein-cond} can be written as the following in this special case:
\begin{equation}
k^2(\phi-\psi) = 12\pi G a^2 (\bar{\rho}+\bar{P})\Sigma +
k^2  (  D_0\phi+\frac{D_1}{k} \dot{\phi} ).
\end{equation}
Ref.~\cite{tessa} then shows that the absence of a term proportional to the metric derivative 
will lead to an inconsistency. However, this conclusion is valid only if one assumes that $\beta$
in Eq. \ref{Agamma} is a function of background quantities, which usually is not the case. 
Otherwise, one can use Eq. \ref{Agamma} to define $\beta$:
\begin{equation}
\beta \equiv \dfrac{12\, \pi G a^2 \, \alpha \Sigma + k^2(D_0\phi+\frac{D_1}{k} \dot{\phi} )}{\psi} ,
\end{equation}
leaving no ambiguity or inconsistency.\footnote{Note that this might be troublesome
if $\psi$ goes to zero at some moments of time. This can happen for the scales that enter the horizon
during radiation domination.}

It is also claimed in Ref.~\cite{tessa2} that the $\{\gamma,\mu\}$ parametrization becomes ambiguous
on large scales, while none of these shortcomings apply to the PPF method. However, these criticisms do 
not seem legitimate, since, as was shown in this section, there is a direct
connection between $\{\gamma$, $\mu\}$, and the PPF parameters.\footnote{In particular there is 
nothing wrong with the $\{\gamma, \mu \}$
parametrization on large scales, since $\psi$ is certainly always non-zero.} For any given 
set of functions for the 
PPF method, one can find a corresponding set of functions $\{\gamma,\mu\}$, using 
Eqs.~\ref{bir} to \ref{dor} and \ref{Agamma}, that will produce the exact 
same result for any observable quantity.  One only needs to ensure the  
use of consistent equations while modifying gravity through codes such as CAMB.

Although we believe that there is no ambiguity in the $\{\gamma,\mu\}$ parameterization,  
we also believe that our $\{A,B,C,D\}$ parameterization can be implemented much more
easily in Boltzmann codes. Furthermore, there is a potential problem for 
the $\{\gamma,\mu\}$ parametrization 
on the small scales that enter the horizon during the radiation domination era. The metric perturbation
$\psi$ will oscillate around zero a couple of times for these scales and that makes the $\gamma$ 
function blind to any modification at those instants of time.
This behaviour also has the potential to lead to numerical instabilities.

\section{Numerical calculation} \label{numeric}
In this section we will constrain the parameterization described in Sec.~\ref{ABparam}
using the CMB anisotropy power spectra. We will describe the effects of the modifying parameters 
on the CMB and show the results of numerical calculations from CAMB and CosmoMC.

\subsection{Effects of $A$ and $B$ on the power spectra}
Before showing numerical results, we first describe some of the physical effects of having 
non-zero values of $A$ or $B$. So far we have not placed any constraints on these quantities, 
which are in general functions of both space and time.
There are some effects that can be explicitly seen from the 
equations of motion and energy conservation. For example, a positive $A$ enhances 
the pressure perturbation and anisotropic stress, while reducing the density perturbation.
On the other hand, a positive $B$ will enhance the momentum perturbation and reduce the 
pressure perturbation and anisotropic stress. 

There are also some other effects that need a little more algebra to see, and we now discuss
three examples.

\textit{Neutrino moments:}

The neutrinos' zeroth and second moments, $\{F_{\nu 0}, F_{\nu 2}\}$, are coupled to the 
modifying gravity terms according to the Boltzmann equations \cite{ma} and  
Eqs.~\ref{first} to \ref{last}:
\begin{eqnarray}
\dot{F}_{\nu 0} &=& - k\, F_{\nu 1}-\dfrac{2}{3} \dot{h} ; \\
\dot{F}_{\nu 2} &=& \dfrac{2}{5} k\, F_{\nu 1} - \dfrac{3}{5} k\,
F_{\nu 3}+ \dfrac{4}{15} ( \dot{h}+6\dot{\eta} ) .
\end{eqnarray}
Here $\dot{h}$ is modified according to Eq.~\ref{first}, and the term 
$\dot{h}+6\dot{\eta}$ is coupled to $A$ and $B$ through Eqs.~\ref{first}
and \ref{second}:
\begin{eqnarray}
\dot{h}+6\dot{\eta} &=& 
\dfrac{2\,k^2\,\eta + 8\, \pi G\,a^2 \delta \rho}{\mathcal{H}}
+ 24 \, \pi G\,a^2 (\bar{\rho} + \bar{p} ) \dfrac{V}{k} \nonumber \\
&& -\dfrac{2\,k^2\,A}{\mathcal{H}} + 6\,k\,B.
\end{eqnarray} 
Therefore, modified gravity can have a significant effect on the 
neutrino second moment.

\textit{Photon moments:}

While the same thing is valid for the photons' second moment after decoupling, the situation is
different during the tight coupling regime. The Thomson scattering rate is so high in the tight
coupling era that it makes the second moment insensitive to gravity. In other words, the 
electromagnetic force is so strong that it does not let the photons feel gravity. 

\textit{ISW effect:}

\begin{figure}[ht]
\includegraphics[scale=0.7]{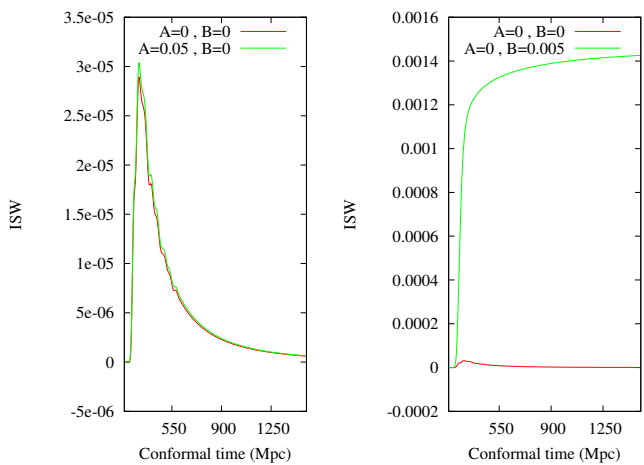}
\caption{\label{fig:ISW} Effects of a constant, non-zero $A$ and
$B$ on the ISW effect. This plot shows the ISW effect for a specific scale of $k=0.21\,
{\rm Mpc^{-1}}$.
} 
\end{figure}

\begin{figure}[ht]
\scalebox{0.7}{\input{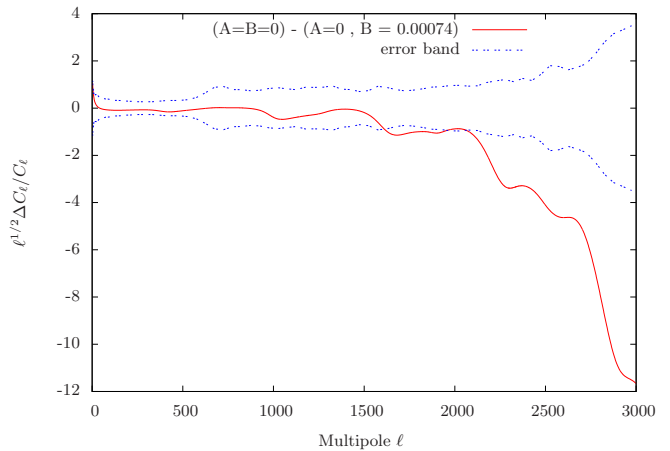}}
\caption{\label{fig:residue} Effects of a constant $B$ on the CMB power spectra.
The plot shows the difference in power for the case of zero $B$ minus the best fit 
non-zero $B$, using \textit{WMAP}9 and SPT12 data, while keeping all the rest of the 
parameters the same.
The error band plotted is based on the reported error on the binned CMB power spectra from
the \textit{WMAP}9 \cite{WMAP9} and SPT12 \cite{SPT12} groups. 
} 
\end{figure}

The integrated Sachs-Wolfe (ISW) \cite{ISW} effect is proportional to $\dot{\phi} + \dot{\psi}$
in the Newtonian gauge. In the synchronous gauge this is
\begin{equation}
\dot{\phi} + \dot{\psi} = \dfrac{\dot{\ddot{h}}+6\, \dot{\ddot{\eta}}}{2\, k^2}
   + \dot{\eta} .
\end{equation}
Here $\dot{\eta}$ is modified according to Eq.~\ref{second} and, therefore, a subtle 
change in the function $B(k,\tau)$, can have a considerable influence on the ISW 
effect. Fig.~\ref{fig:ISW} shows the effects of a constant non-zero $A$ and $B$ on 
the ISW effect. For the case of a constant non-zero $B$, the ISW effect is always 
present, since the time derivative of the potential is constantly sourced by 
this function. This will result in more power on all scales, including the tail of 
the CMB power spectrum (see Fig.~\ref{fig:residue}). Fig.~\ref{fig:residue} shows 
that if a non-zero $B$ is favoured by CMB data, it will be mostly due to the large 
${\ell}$s, ($\ell > 1500$), and comes from its anti-damping behaviour.

\begin{figure}[h]
\scalebox{0.7}{\input{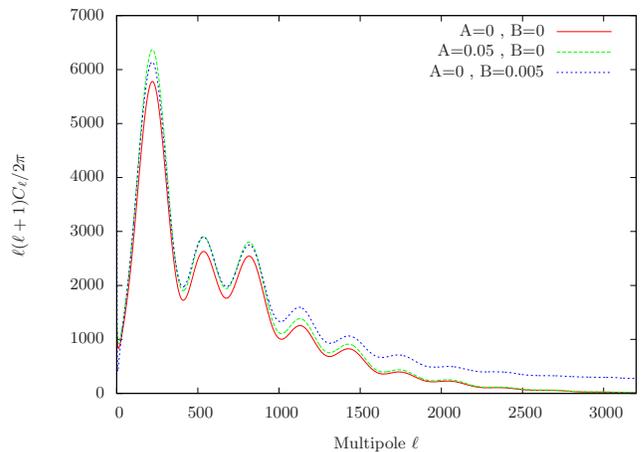}}
\caption{\label{fig:camb} Effects of a constant, non-zero $A$ or
$B$ on the CMB power spectra. One can see that the two parameters 
have quite different effects.} 
\end{figure}

Fig.~\ref{fig:camb} shows the CMB power spectra for the case of a constant but non-zero $A$ or $B$.
Note how a constant non-zero $B$ raises the tail of the spectrum up. One
might also point to the degeneracy of $A$ and the initial amplitude (usually called $A_{\rm s}$)
mostly by looking at the height of the peaks. 

\textit{Matter overdensity:}

The Boltzmann equation for cold dark matter overdensity in the synchronous gauge reads \cite{ma}
\begin{equation}
\dot{\delta}_{\rm CDM} \equiv \left(\dfrac{{\delta \rho}_{\rm CDM}}{\bar{\rho}_{\rm CDM}} \right)^{\textbf{.}} = 
- \dfrac{1}{2} \, \dot{h} .
\end{equation} 
Using Eqs.~\ref{first}, \ref{second}, and the Friedmann equation, 
$\mathcal{H}^2 = \dfrac{8\, \pi\, G\, a^2}{3} \bar{\rho}$, and assuming a matter-dominated 
Universe with no baryons, one obtains the following equation for the cold dark matter overdensity:
\begin{eqnarray}
&& \mathcal{H} \, \dot{\delta}_{\rm CDM} =  -\dfrac{3\,\mathcal{H}^2}{2} {\delta}_{\rm CDM}
+ k^2 A - k^2 \eta,  \nonumber \\ \mbox{i.e.}
&& \ddot{\delta}_{\rm CDM} + \left(\dfrac{\dot{\mathcal{H}}}{\mathcal{H}} + \dfrac{3}{2} \mathcal{H}
\right) \dot{\delta}_{\rm CDM} + 3\,\dot{\mathcal{H}} {\delta}_{\rm CDM} = \nonumber \\
&& \qquad \quad - \dfrac{k^3}{\mathcal{H}} B
+ \dfrac{k^2}{\mathcal{H}} \dot{A}  \label{matter} .
\end{eqnarray}
The above equation clearly shows the role of $A$ and $B$ as driving forces for the 
matter overdensity. The $k^3$ prefactor makes the first term on the right hand side 
dominant on small scales and this will therefore have a significant effect on the 
matter fluctuation amplitude at late times. The matter power spectrum will therefore 
be expected to put strong constraints on modified gravity models.

\subsection{Markov Chain constraints on $A$ and $B$}
Since $A(k,\tau)$ and $B(k,\tau)$ are free functions, we need to choose some simple cases
to investigate. We choose here to focus on the simple cases of $A$ and $B$ being separate 
constants (i.e. independent of both scale and time). We do not claim that this is in any 
sense a preferred choice --- we simply have to pick something tractable.
With better data one can imagine constraining a larger set of parameters, for example 
describing $A$ and $B$ as piecewise constants or polynomial functions.

\begin{figure}[h]
\centering
\scalebox{0.45}{\input{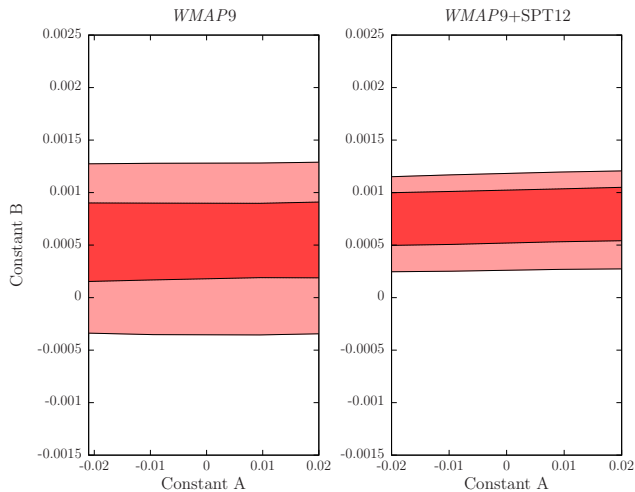}}
\caption{\label{fig:ab_cons} $68$ and $95$ percent
contours of the constants $A$ and $B$ using \textit{WMAP}9-year data alone
(left) and SPT12 (right) without including lensing effects, and neglecting
late time growth effects.} 
\end{figure}

We used CosmoMC to constrain constant $A$ and $B$, together with the \textit{WMAP}-9 
\cite{WMAP9} and SPT$12$ \cite{SPT12} CMB data. The amplitudes of the CMB foregrounds were 
added as additional parameters and were marginalized over for the case of SPT$12$. The 
resulting constraints and distributions are shown in Fig.~\ref{fig:ab_cons}.
Here we focus entirely on the effects of $A$ and $B$ on the CMB. Hence we turn off the 
post-processing effects of lensing~\cite{lensing}, and ignore constraints from any other astrophysical
data-sets.

\begin{figure}[h]
\centering
\scalebox{0.45}{\input{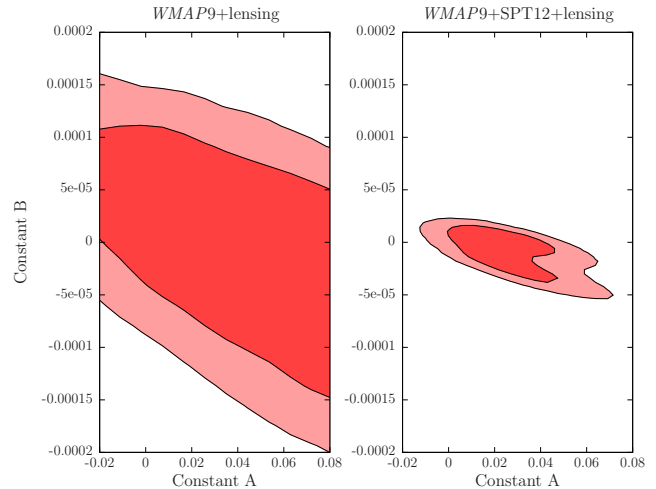}}
\caption{\label{fig:lensing} $68$ and $95$ percent
contours of the constants $A$ and $B$ using \textit{WMAP}9-year data alone
(left) and \textit{WAMP}9 $+$ SPT12 (right), with lensing effects included.} 
\end{figure}

One might conclude from Fig.~\ref{fig:ab_cons} that general relativity is ruled out
by nearly $3\sigma$ using CMB alone, since a non-zero value of $B$ is preferred. 
However, adding lensing to the picture will considerably
change the results. As was shown in Eq.~\ref{matter}, a non-zero $B$ will change 
the matter power spectrum so drastically that in a universe with non-zero $B$, lensing 
will be one of the main secondary effects on the CMB. The results of a Markov chain 
calculation that includes the effects of lensing (i.e. assuming $B$ was constant
not only in the CMB era, but all the way until today)
is shown in Fig.~\ref{fig:lensing}, and are entirely consistent with GR.

\begin{figure}[h]
\centering
\scalebox{0.45}{\input{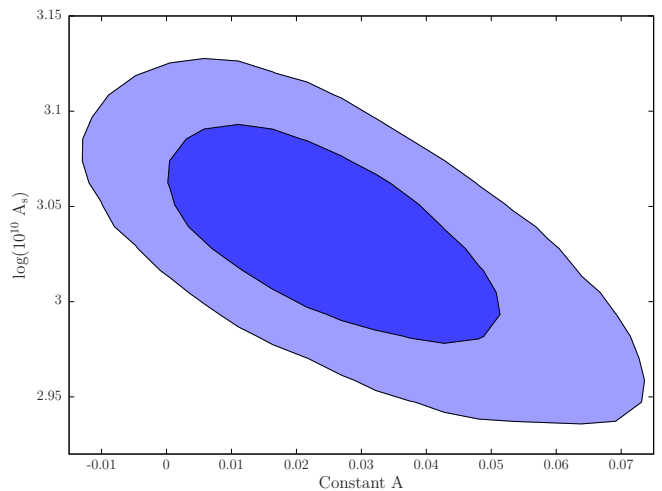}}
\caption{\label{fig:A_A_s} The strong anti-correlation between parameter
$A$ and the initial amplitude, $A_{\rm s}$, makes the constraints on either one 
of these two parameters weaker.} 
\end{figure}

The broad constraint on $A$ is mainly due to a strong (anti-)correlation between 
$A$ and the initial amplitude of the scalar perturbations. 
Two-dimensional contour plots 
of $A$ versus $A_{\rm s}$, the initial 
amplitude of scalar curvature perturbations, are shown in Fig.~\ref{fig:A_A_s}.

The mean of the likelihood and $68\%$ confidence interval for the six cosmological parameters
together with $A$ and $B$ are tabulated in Table \ref{table:dist}. Note that the simplified 
case we are considering here treats CMB constraints only. If we really took $B$ as constant 
for all time, then there would be large effects on the late time growth, affecting the matter 
power spectrum, and hence tight constraints coming from a relevant observable, such as $\sigma_8$ 
from cluster abundance today.

\begin{table*}
\nointerlineskip
\setbox\tablebox=\vbox{
\newdimen\digitwidth 
\setbox0=\hbox{\rm 0} 
\digitwidth=\wd0 
\catcode`*=\active 
\def*{\kern\digitwidth} 
\newdimen\signwidth 
\setbox0=\hbox{{\rm +}} 
\signwidth=\wd0 
\catcode`!=\active 
\def!{\kern\signwidth} 
\newdimen\pointwidth 
\setbox0=\hbox{\rm .} 
\pointwidth=\wd0 
\catcode`?=\active 
\def?{\kern\pointwidth} 
\halign{\hbox to 1.75in{#\leaderfil} &
 &
\hfil#\hfil\tabskip=0.0em\cr
\noalign{\doubleline}
Parameter & WMAP9 & $\mbox{WMAP9+SPT12}$ & $\mbox{WMAP9+SPT12+lensing}$\cr
\noalign{\vskip 3pt\hrule\vskip 4pt} 
$100\, \Omega_{\rm b} h^2$& $2.22 \pm 0.05$& $2.10 \pm 0.03$ & $2.22 \pm 0.04$\cr
$\Omega_{\rm DM} h^2$& $0.118 \pm 0.005$& $0.122 \pm 0.005$ &$0.115 \pm 0.004$\cr
$100\, \theta$& $1.038 \pm 0.002$& $1.040 \pm 0.001$ &* $1.042 \pm 0.0010$ \cr
$\tau$&  $ 0.086 \pm 0.014$& $0.076 \pm 0.012$ & $0.084 \pm 0.012$\cr
${\rm log}(10^{10} A_{\rm s})$ & $3.1 \pm 0.2$& $3.1 \pm 0.2$ & $3.0 \pm 0.4$\cr
$n_{\rm s}$& $0.96 \pm 0.01$& $0.93 \pm 0.01$ & $0.96 \pm 0.01$\cr
$100 A$&?*$1 \pm 10$?& $0.02 \pm 10$ * & $2.7 \pm 1.7$\cr
$1000 B$&$0.44 \pm 0.41$& $0.74 \pm 0.24$ &$-0.0097 \pm 0.016$*!\cr
\noalign{\vskip 3pt\hrule\vskip 4pt}
}}
\endtable
\caption{The mean likelihood values together with the
    $68\%$ confidence interval for the usual six cosmological parameters, together 
    with constant $A$ and $B$, using CMB constraints only.}
    \label{table:dist}
\end{table*}

\subsection{Alternative powers of $k$ in $B$}
\label{Sec:BH_K}
\begin{figure}[h]
\centering
\scalebox{0.67}{\input{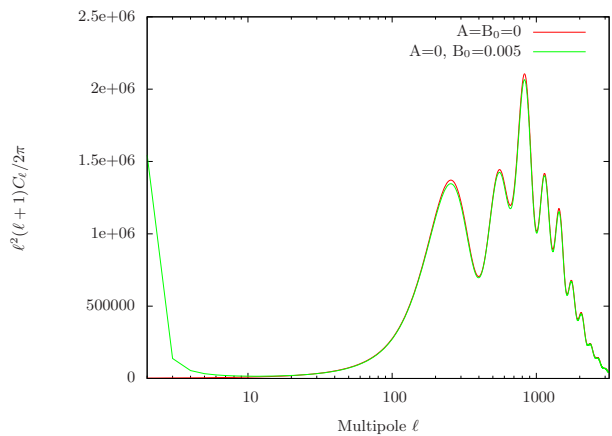}}
\caption{\label{fig:BH_K} 
Effect of a non-zero $B_0$ on the CMB power spectra, with the choice
$B = \mathcal{H}\,B_0/k$.} 
\end{figure}

Examining Eq.~\ref{matter} reveals that the only term modifying the matter
power spectrum in the case of constant $A$ and $B$, is $k^3B/ \mathcal{H}$.
This term is important for two reasons. Firstly, this is the only term
introducing a $k$ dependence in the cold dark matter amplitude at late times
and at sufficiently large scales where one can completely ignore the effect of baryons
on the matter power spectra. Secondly, the $k^3$ factor
enhances this term significantly on small scales in the case of a constant $B$.
Since the amplitude of matter power spectrum (via lensing effects) was 
the main source of constraints on $B$,
it would seem reasonable to choose $B = \mathcal{H}\,B_0/k$, where $B_0$ is a 
dimensionless constant. This should avoid too much power in the matter densities on
small scales, and therefore reduce lensing as well. 
However, this choice will lead to enormous power on 
the largest scales, as is shown in Fig.~\ref{fig:BH_K}.

In order to match with data, one could choose a form in which 
$B$ switches from $B = \mathcal{H}\,B_0/k$ to 
$B = \mathcal{H}\,B_0/k_0$, where $k_0$ is some small enough transition scale.
We discuss this simply as an alternative to the $B = constant$ case. There is clearly 
scope for exploring a wider class of forms for the functions $A(k,\tau)$ and
$B(k,\tau)$.

\section{Discussion} \label{discuss}
Since a constant $A$ is essentially degenerate with the initial amplitude of the
primordial fluctuations, the CMB alone cannot constrain this parameter.
On the other hand, constant $B$ seems to be fairly well constrained by the CMB data. 
However, if $B$ was an oscillating function of time, changing sign from time to time, 
its total effect on the CMB power spectra would become weaker and the constraints would 
be broader. According to Eq. \ref{second}, a constant $B$ will change $\eta$ monotonically, 
while the effect of an oscillating $B$ will partially cancel some of the time. Together, 
the results of Sec.~\ref{numeric} show that the ISW effect and the growth at relatively 
recent times (driving the amplitude of matter perturbations)
can have huge constraining power for many generic theories of modified gravity.
(See Ref.~\cite{galile} for a recent example).
One can consider different positive or negative powers of ($\mathcal{H}/k$) as part of the
dependence of $B$ in order to get around the matter constraints, as was discussed
in Sec.~\ref{Sec:BH_K}. 

We have seen that when considering CMB data alone, there seems to be a mild preference for
non-zero $B$. This is essentially because it provides an extra degree of freedom for 
resolving a mild tension between \textit{WMAP} and SPT. Neveretheless it remains true that
a model with $B$ constant for all time would be tightly constrained by observations of the matter
power spectrum at redshift zero. We leave for a future study the question of whether there might
be any preference for more general forms for $A(k,\tau)$ and $B(k,\tau)$ using a combination of
\textit{Planck} CMB data and other astrophysical data-sets.

\section*{Acknowledgments}
We acknowledge many detailed and helpful discussions with our colleague James Zibin. 
We also had very useful conversations about this and related work with Levon Pogosian. 
This research was supported by the Natural Sciences and Engineering Research Council of Canada.

\newpage
\bibliographystyle{unsrt}
\bibliography{mog}

\end{document}